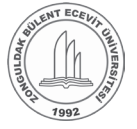

# Design of Non-Inverting Buck-Boost Converter for Electronic Ballast Compatible with LED Drivers

*LED Sürücülerle Uyumlu Elektronik Balastlar için Evirmeyen Alçaltıcı-Yükseltici Dönüştürücü Tasarımı*

Ridvan Keskin[*] ⬥, Ibrahim Aliskan ⬥

Zonguldak Bulent Ecevit University, Faculty of Engineering, Department of Electrical and Electronics Engineering, Zonguldak, Turkey

## Abstract

This paper presents design and control of dual-switch non-inverting buck-boost converter (CBB). This converter is designed to simplify the compatibility of electronic ballast with simple and low cost LED drivers. The converter provides starting voltage and current limitation of electronic ballasts, which operates at continuous conduction mode (CCM). The voltage of load terminal is controlled by adjusting the duty cycle of the PWM regulator. Although both converter switches are controlled separately, one feedback control loop is needed to obtain the desired compensator level. Appropriate control requirements have been defined by analyzing open-loop characteristic of converter transfer function through the small-signal model of CBB, which lets decide about the control strategy and analyse the stability and performance of the closed loop control system. In order to obtain the desired output voltage, Type-III rational controller is preferred because of the non-minimum phase feature in the converter boost mode. The performance of the synthesized voltage controller is verified by comparing of the pre-determined performance requirements and the obtained simulation results.

## Öz

Bu çalışmada iki anahtarlı evirmeyen alçaltıcı-yükseltici dönüştürücü tasarımı ve kontrolü sunulmaktadır. Sürekli iletim modunda çalışan dönüştürücü, balastların temel işlevi olan ateşleme voltajını ve akım sınırlamasını sağlamaktadır. Bu dönüştürücü basit ve düşük maliyetli LED sürücüleri ile elektronik balastların uyumunu basitleştirmek için tasarlanmıştır. Dönüştürücünün anahtarları ayrı ayrı kontrol edilmesine rağmen istenen kompansatör seviyesini elde etmek için bir geri besleme kontrol döngüsü kullanılmıştır. Uygun kontrol gereksinimleri dönüştürücünün küçük sinyal modeli ve sistemin transfer fonksiyonunun açık döngü karakteristiği analiz edilerek tasarlanmıştır. Arzulanan çıkış gerilimini elde etmek için Tip-III rasyonel kompansatör tercih edilir. Çünkü dönüştürücü yükseltme modunda minimum fazlı olmayan bir sistemdir. Kontrolcünün performansı, gerçek zamanlı performans gereksinimleri ve elde edilen simülasyon sonuçları karşılaştırılarak doğrulanmıştır.

**Anahtar Kelimeler:** Tip-III kontrolcü, Küçük-sinyal modeli, Elektronik balast, LED sürücü

## 1. Introduction

Incandescent lamps and fluorescent lamps have been utilized in residential and industrial sectors in 20th century. This lamps have low lumens per unit in comparison with power level they consume. Nowadays, high efficiency fluorescent lamps and light-emitting diodes (LED) have emerged thanks to technological developments and efficiency enhancement efforts. However, LEDs and fluorescent lamps do not connect directly to the ac mains. Fluorescent lamps initially require high voltage and current limitation after ignition. LED lamps also require current limiting and must be operated at the appropriate voltage level according to the type to be used. Ballasts and LED drivers fulfill these requirements. A ballast has two primary functions. First is to generate ignition voltage, and the second is to regulate fluorescent lamp current because the lamp has a $V_L$-$I_L$ characteristic with a negative slope, resulting in a negative dynamic resistance $R_L = dV_L/dI_L$ (Chondrakis and Topali 2009). There are a lot of studies about the design of ballast and led drivers (Galkin et al. 2012, Choi and Lee 2012, Cheng et al. 2001, Istók 2015, Ahmed et al. 2015).

―――――――――――――――
*Corresponding author: ridvan.keskin@beun.edu.tr

Ridvan Keskin ⬥ orcid.org/0000-0002-3232-0928
Ibrahim Aliskan ⬥ orcid.org/0000-0003-3901-4955

LEDs have significant advantages in comparison with fluorescent lamps such as long lamp life, low power consumption, containing no mercury and luminous quality. Although LEDs are superior to fluorescent lamps in terms of such advantages, driver circuits of LEDs have disadvantages due to their cost and inadequacy in the market (Choi et al. 2015, Liang et al. 2013). Therefore, the operation of LED lamps with electronic ballast, which is a ballast type having high input power factor, low input current harmonics, good lamp current crest factor and simply replaces fluorescent lamps with LED lamps, raise the issue.

In (Chen and Chung 2011), A driving technique that could operate with electronic ballast for fluorescent lamps to drive LEDs without extra component is presented. It provides a solution to turn a lighting system with fluorescent lamps into the one with LED lamps. An LED lamp driver compatible with electronic ballasts is designed (Chen and Chung 2013). The driver could be powered with low-frequency electromagnetic ballasts and by the ac mains. In addition, this driver acts as a phase shift resonant converter when operating with a ballast. In (Shao and Stamm 2013), one of the simplest ways of operating LED lamps with electronic ballast is presented. Here, only passive rectifier is added between a LED lamp and the ballast. However, this simple design can't adapt the LED lamps operating at different voltage levels. Furthermore, since the ballasts produced by various companies have different constant current outputs, a design is needed to solve this issues. In (Shao and Stamm 2016), a solution to this issue has been produced, but the autotransformer and current-fed buck converter used for voltage balancing has created additional cost and redundant components. Neither an inductor nor an electrolytic capacitor is introduced in the proposed LED driver (Lee et al. 2016). It is compatible with different ballast types such as instant start, rapid start, and programmed start types of the electronic ballasts. However, the operation of LED lamps at different voltage levels according to fluorescent lamps still remains a problem. Therefore, ballast circuit designs that could adapt to different voltage levels and have low cost are needed, where the fluorescent lamp is replaced with an LED lamp.

Figure 1 shows the block diagram of a two-stage electronic ballast with PFC and connection a LED Driver. The ballast consists of EMI filter, power factor corrector, high-frequency dc/ac inverter, and control circuitry. PFC consists of an active or passive power factor correction circuit such as a valley-circuit and boost converter. In order to increase the power factor at the desired level (PF> 0.98), active power factor correction circuits are preferred. Usually boost converter which is operated in boundary conduction mode (BCM), Discontinuous Conduction Mode (DCM) is used as active PFC because of simple implementation and low cost. Half Bridge or Full Bridge Inverter is used for dc/ac conversion (Nguyen et al. 2015). Finally, the resonant tank consists of simple L-C resonant circuits.

Figure 2. shows the control of a non-inverting buck-boost converter. The converter consists of the two active switches, two diodes acts as passive switches and a L-C low-pass filter. Resistance $r_L$ represents the equivalent series resistance (ESR) of inductor L. In practice, the series equivalent resistance $r_C$ of capacitor C is usually very low but it is not neglected in this analysis due to significant value in frequency domain analysis. Transistors could be used instead of diodes at low power levels to reduce the voltage drop caused by the diodes because MOSFETs have a very low resistance on state (Callegaro et al. 2017). The system is capable of converting the supply voltage source to higher and lower voltages to

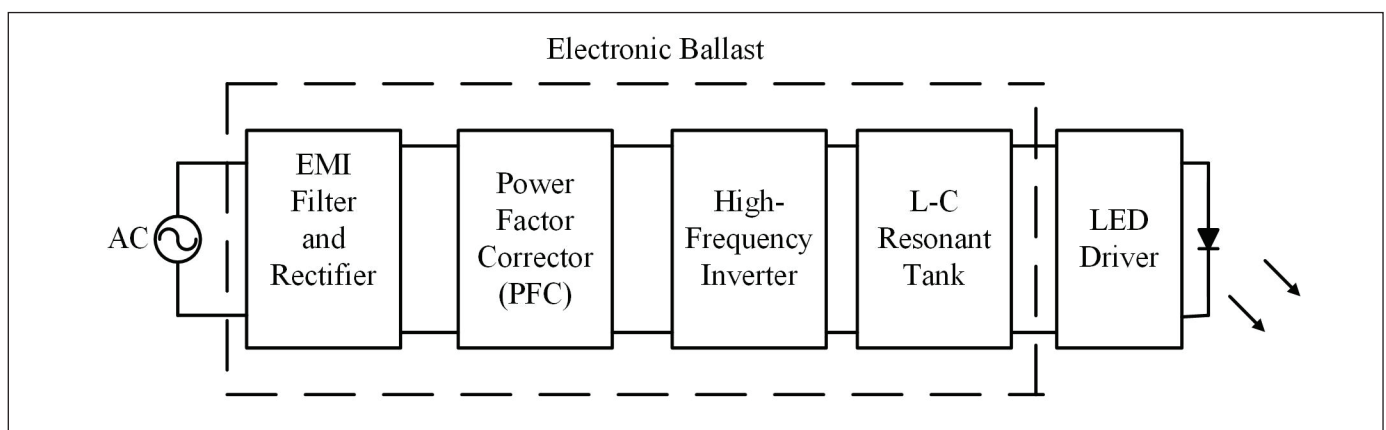

**Figure 1.** Electronic ballast circuit compatible with a LED driver.

the load terminal with voltage polarity unchanged. The converter is used different input and output voltage-range requirements such as battery charge, photovoltaic power system, PFC applications thanks to this feature (Lu and Nguyen 2012, Ugale and Dixit 2017). Besides, the converter has simple structure, low stress on switches and positive polarity of output voltage in comparison with other buck-boost topologies such as SEPIC, Cuk, one switch buck-boost converter (He et al. 2013).

The buck-boost converter transfer function is different for buck and boost modes. In the transfer function of the converter boost mode there is a zero in the right half plane. As well as uncertainties in the system model, material life, aging and noise make controller synthesis difficult. Type-III rational controller structure is used to provide performance requirements for both buck mode and boost mode under uncertainties in the system. Because of the rational selection of the synthesized voltage controller it could be implemented with appropriate op-amp circuits. The schematic diagram of the close-loop buck-boost converter is given in Figure 2.

In this paper, the design and closed-loop control of a non-inverting buck-boost converter are presented. The converter is used for two main functions. First, it increases the power factor of the electronic ballast. Second, it produces starting voltage to ignite the fluorescent lamps at starting. This will make it easier to operate ballasts with LED driver circuits operating at different voltage levels. The purpose of this paper is to present design equations of non-inverting buck-boost circuit, a frequency domain steady-state analysis, controller design and to verify all this steps with simulation results.

The paper is divided in six chapters. Chapter two includes to design the converter, occuring modes of the converter and producing gate signals of the converter. Transfer function of the converter, only control input is duty cycle, is obtained in chapter three. In chapter four, frequency response of the converter and required performance characteristics are presented. In chapter five, simulations results are presented. In conclusion, discussions and considerations of the paper is given.

## 2. Material and Methods

Consider non-inverting buck-boost circuit which consists of two independent active power switches ($SW_1$, $SW_2$) are driven by two PWM signals of $PWM_1$ and $PWM_2$, respectively. In this circuit, there are four different operating modes according to the operating states of the switches. These modes are presented in Table 1.

### 2.1. Boost Mode

If input voltage is less than the terminal voltage of the fluorescent lamp, boost mode of operation is used to step-up the input voltage to ignite the lamp. In this mode, the power switch $SW_1$ is always on, while the power switch $SW_2$ is operated with duty cycle $D_2$. In comparison with boost converter, this mode has a few drawbacks. The drawbacks are voltage drops due to D2 diode and using $SW_1$ MOSFET. But, the voltage drops is ignorable in this paper because of high operating voltage levels. The duty cycle of boost mode is given in Equation (1).

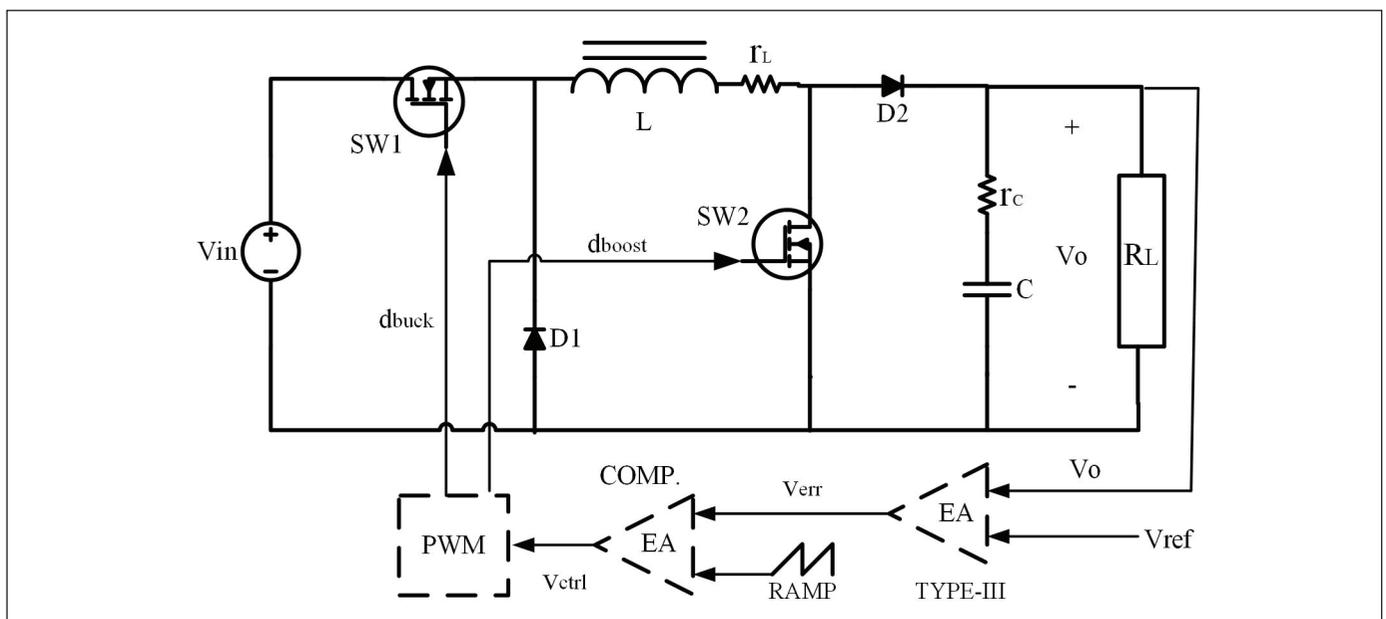

**Figure 2.** Simplified block diagram of non-inverting buck-boost converter.

**Table 1.** Switching states of non-inverting buck boost converter.

| Switching State | SW1 | SW2 | Mode |
|---|---|---|---|
| 1 | ON | PWM | Boost |
| 2 | PWM | PWM | Buck-Boost |
| 3 | OFF | PWM | N/A |
| 4 | PWM | OFF | Buck |

**Table 2.** Circuit parameters

| Parameter | Value | Unit |
|---|---|---|
| Input Voltage | 310 | V |
| Output voltage range | 280-400 | V |
| Power of the lamp | 18 | W |
| Operating frequency | 40 | kHz |
| Inductor value | 15 | mH |
| Capacitor value | 1 | uF |
| ESR of The inductor | 100 | mΩ |
| ESR of The Capacitor | 50 | mΩ |

$$\frac{V_0}{V_{in}} = \frac{1}{1-D_2} \tag{1}$$

### 2.2. Buck Mode

If input voltage is greater than terminal voltage of the fluorescent lamp, buck mode is used. The power switch $SW_1$ is operated with duty cycle $D_1$, while the power switch $SW_2$ is always off. By comparison with buck converter, this mode has one disadvantage which is voltage drops owing to D2 diode. Buck and boost modes analyzed with details (Badawy et al. 2016). The duty cycle of buck mode is given in Equation (2).

$$\frac{V_0}{V_{in}} = D_1 \tag{2}$$

The third mode is prevented because the mode never occurs in buck or boost mode of the converter. Buck-boost mode is used when output voltage almost equal to input voltage. This operation mode is avoided due to high losses of working in this mode in comparison with buck and boost modes (Schaltz et al. 2008). Inductor value and output filter capacitor value are same in all modes of operation of the converter. The values are calculated considering the converter operates in CCM. Equations (3), (4) show the incremental inductor-current during the on time in the buck and boost mode, respectively. It is clear that the instantaneous values of incremental inductor-current for the buck and boost mode are different. Therefore, the different instantaneous inductor-currents at the transitions between the buck and boost mode cause the distortions on the inductor current. But, transitions of the modes are out of scope because of fluorescent lamp loading conditions in this work (Chondrakis and Topalis 2009).

$$\Delta I_L = \frac{V_{in} - V_0}{L} \cdot T_{on} \tag{3}$$

$$\Delta I_L = \frac{V_{in}}{L} \cdot T_{on} \tag{4}$$

Operating parameters of the converter is shown in Table 2. The switching frequency is chosen as 40 kHz because of two reasons. First reason is real-time application considerations. Second reason is to decrease the switching losses. In addition, equivalent resistance value of the fluorescent lamp is high at starting. So, this causes that the value of the inductor is high because the inductor value directly proportional with the resistance value. But it can be decreased significantly by choosing higher switching frequency or reducing upper voltage value of the boost mode.

### 2.3. Pulse Width Modulation

Producing gate signals for controlling the converter switches with Pulse Width Modulation (PWM) is shown in Figure 3. It is clear that PWM pulses are generated by intersection of a control signal (Vctrl) and one carrier signals to avoid the occurrence of converter buck-boost mode. The triangular wave with the maximum amplitude ($V_{H1}$) and minimum amplitude ($V_{L1}$) with two reference line determine the state of $S_1$ and $S_2$, respectively. $G_1$ and $G_2$ are PWM signals driving $S_1$ and $S_2$ switches.

### 3. System Modelling

Power stages of PWM converters are pretty nonlinear systems owing to the fact that they contain at least two switches as one transistor and one diode. To calculation of the characteristics of the converter, nonlinear power stages of the PWM converter should be averaged and linearized. Circuit averaging method and state-space averaging method are two averaging methods for PWM converters. However, the state-space averaging method needs to a considerable amount of differential equations and matrix algebra manipulations. In addition, if the equations has a number of parasitic components, the method is very complex and tedious. That's why, circuit averaging method is used to obtain averaged and linearized model of the converter. Suppose that the frequencies of variations of the converter inputs are much lower than the switching frequency, then the small signal averaged model is a valid representation of converter performance in response to small AC variations about the equilibrium operating point (Kazimierczuk 2015). Therefore, the small-signal model of the converter is presented in Figure 4.

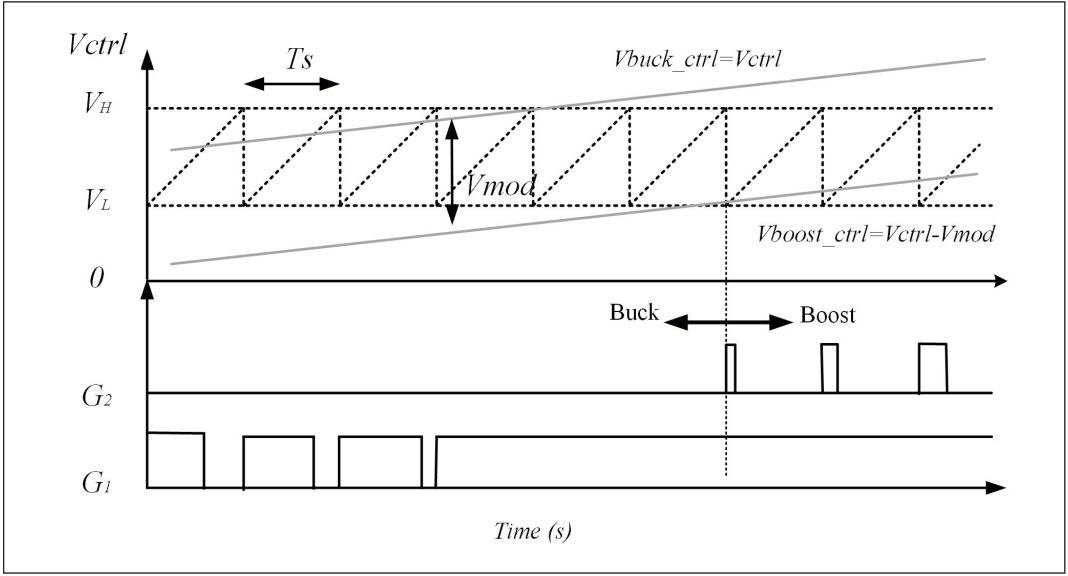

**Figure 3.** PWM modulation strategy.

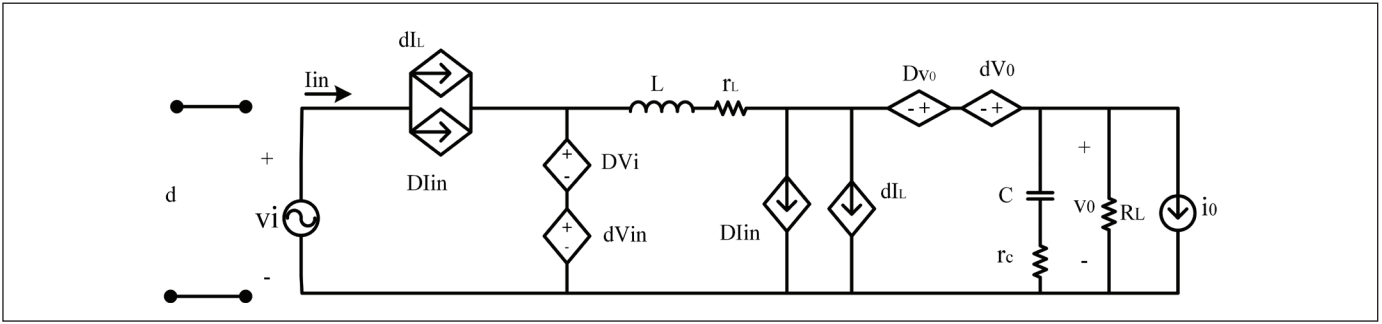

**Figure 4.** Small-signal model of the PWM non-inverting buck-boost converter for CCM operation

The model has two parasitic components. They are inductor resistance and capacitor resistance denoted by $r_L$ and $r_C$, respectively. Other parasitic components are ignored because effect of them to the circuit approximately equal to zero. It is assumed that input voltage of the converter is constant because converter connects directly the ac main. In Figure 4, it is clearly seen that the small-signal model has three input variables $d$, $v_i$, and $i_0$, and one output variable $v_o$. The small-signal duty cycle $d$ is a control variable, while small-signal input voltage $v_i$ and the small-signal load current $i_0$, $v_0$ are disturbances. Firstly, control to output transfer functions for the power stages of the converter are obtained as follows;

$$Z_1(s) = r_L + sL \tag{5}$$

$$Z_2(s) = \frac{R_L \cdot \left(r_c + \frac{1}{sC}\right)}{R_L + r_c + \frac{1}{sC}} \tag{6}$$

$Z_1$ and $Z_2$ are impedances of the low pass filter that consist of the iron-core coil, the capacitor and load presented in Figure 2. From voltage divider, Equation (7) is obtained for buck mode;

$$V_0(s) = Vind(s) \cdot \frac{Z_2(s)}{Z_1(s) + Z_2(s)} \tag{7}$$

Supposing that $v_i$ and $i_0$ equal to zero. Substitution of Equations (5) and (6) into Equation (7) yields the transfer function of buck converter denoted by K1(s) in s-domain.

$$K_1(s) = \frac{v_0(s)}{d(s)} = \frac{V_{in}R_L r_C}{L(R_L + r_C)} \cdot A$$

$$A = \frac{s + \frac{1}{Cr_c}}{s^2 + s \cdot \frac{C(R_L r_c + R_L r_L + r_c r_L) + L}{LC(R_L + r_c)} + \frac{R_L + r_L}{LC(R_L + r_c)}}$$

(8)

In boost mode, consider that the current through $Z_2$ impedance is $i_{Z_2}$ the current through the inductor is $I_{in}$. According to the Kirchoff Current Law, $I_{in}$ is

$$I_{in} = D_2 I_{in} + I_L d + i_{Z_2}$$

(9)

from Equation (9), produces

$$I_{in} = \frac{I_L d}{1 - D_2} + \frac{v_0}{(1 - D_2) Z_2} = \frac{V_0 d}{(1 - D_2)^2 R_L} + \frac{v_0}{(1 - D_2) Z_2}$$

(10)

Using the Kirchoff Voltage Law,

$$D_2 v_0 + V_0 d = I_{in} Z_1 + v_0$$

(11)

Substituting Equation (10) into Equation (11) yields

$$V_0 d = v_0 (1 - D_2) + \frac{V_0 Z_1 d}{R_L (1 - D_2)^2} + \frac{v_0 Z_1}{Z_2 (1 - D_2)}$$

(12)

Finally, which becomes

$$v_0 (1 - D_2) \left[ 1 + \frac{Z_1}{(1 - D_2)^2 Z_2} \right] = d V_0 \left[ 1 - \frac{Z_1}{(1 - D_2)^2 R_L} \right]$$

(13)

Hence, substitution of (5) and (6) into (13) gives the control to output transfer function of the boost mode in the s-domain.

$$K_2(s) = \frac{V_0 r_c}{(D_2 - 1)(R_L + r_c)} \frac{(s + \frac{1}{Cr_c}) \cdot Z}{s^2 + sX + Y}$$

$$Z = \left( s - \frac{1}{L} [R_L (1 - D_2)^2 - r_L] \right)$$

$$X = \frac{r_L + (1 - D_2)^2 R_L}{LC(R_L + r_c)}$$

(14)

$$Y = \frac{C(r_L (R_L + r_c) + (1 - D_2)^2 R_L r_c) + L}{LC(R_L + r_c)}$$

where load resistance, inductor resistance, capacitor resistance, duty cycle of the boost mode and maximum dc output value of the boost mode are denoted by $R_L$, $r_L$, $r_c$, $D_2$, $V_0$, respectively. In addition that, $I_L$ is averaged current through the inductor. The control to output transfer functions denoted by $K_1(s)$, $K_2(s)$ are also termed duty ratio to output transfer function. Because only control input is duty ratios of $SW_1$ and $SW_2$ switches.

## 4. Controller Synthesis

The compensator could be designed in two ways as one level or two level. One level is the design which selection as regards worse open loop frequency response in the bode diagram. Two level control is that compensator design for buck and boost modes separately. This way requires redundant processing chunks and requires an extra loop for microprocessors. Therefore, a single-level classical controller will be designed.

Frequency response of buck and boost modes is presented in Figure 5. It is obvious that the converter is stable in buck mode in Figure 5. By designing a compensator with a wide bandwidth, this mode can be brought to the desired level. However, the boost mode contains right half plane zero (RHP), which limits the crossover frequency of the closed-loop non-inverting buck-boost converter, where the position depends on the inductor value and the inductor current. In other words, there is a non-minimum phase feature in the boost mode. It is clear that the phase response of the mode is worse than the buck mode in Figure 5. These reasons decrease the stability of boost mode in comparison with the buck mode. Therefore, the single level compensator will be designed in reference to the boost mode. Since the operating frequency is 40 kHz, the cut-off frequency should be selected as 1-10 kHz in view of implementation concerns. Control flow diagram of the system is presented in Figure 6. The figure shows producing duty cycles of buck and boost modes to supply desired output voltage.

Parametric transfer function of the Type-III rational controller is given in Equation (15). The controller adds a pole at s = 0 and eliminates the complex poles that retard the system.

$$K(s) = \frac{a_1 s^2 + a_2 s + a_3}{s \cdot (b_1 s^2 + b_2 s + b_3)}$$

(15)

The goals or design criterias determined for the converter are given in Table 3. The design criterias are determined considering real-time performance, by analyzing open-loop characteristic of converter transfer function and the recommendations in (Kazimierczuk 2015). The controller coefficients are calculated by determining the closed-loop poles corresponding to this performance requirement. Obtained controller coefficients are given in Table 4.

## 5. Discussion and Simulation Results

Various 18 W fluorescent lamps are operated in the laboratory with appropriate electronic ballasts. The initial voltage of the fluorescent lamps is measured to be about 340 volts and over. Furthermore, it is measured that the operating DC link voltage of the ballasts is about 280 volts. Considering the operating conditions of the fluorescent lamps in (Chondrakis and Topalis 2009) and the measured values, it is decided that reference values of the converter are 400 V and 280 V for boost and buck modes, respectively.

**Table 3.** Performance goals of the non-inverting buck-boost converter

| Parameter | Value |
|---|---|
| Rise time [$t_r$] | < 0.1 ms |
| Settling time [$t_s$] | < 0.25 ms |
| Steady state error [$e_{ss}$] | 0 |
| Overshoot [$M_p$] | < %15 |

**Table 4.** Controller coefficients.

| Parameter | Value |
|---|---|
| $a_1$ | 1.9e-6 |
| $a_2$ | 0.012915 |
| $a_3$ | 80 |
| $b_1$ | 6.8e-12 |
| $b_2$ | 0.0000030 |
| $b_3$ | 1.5 |

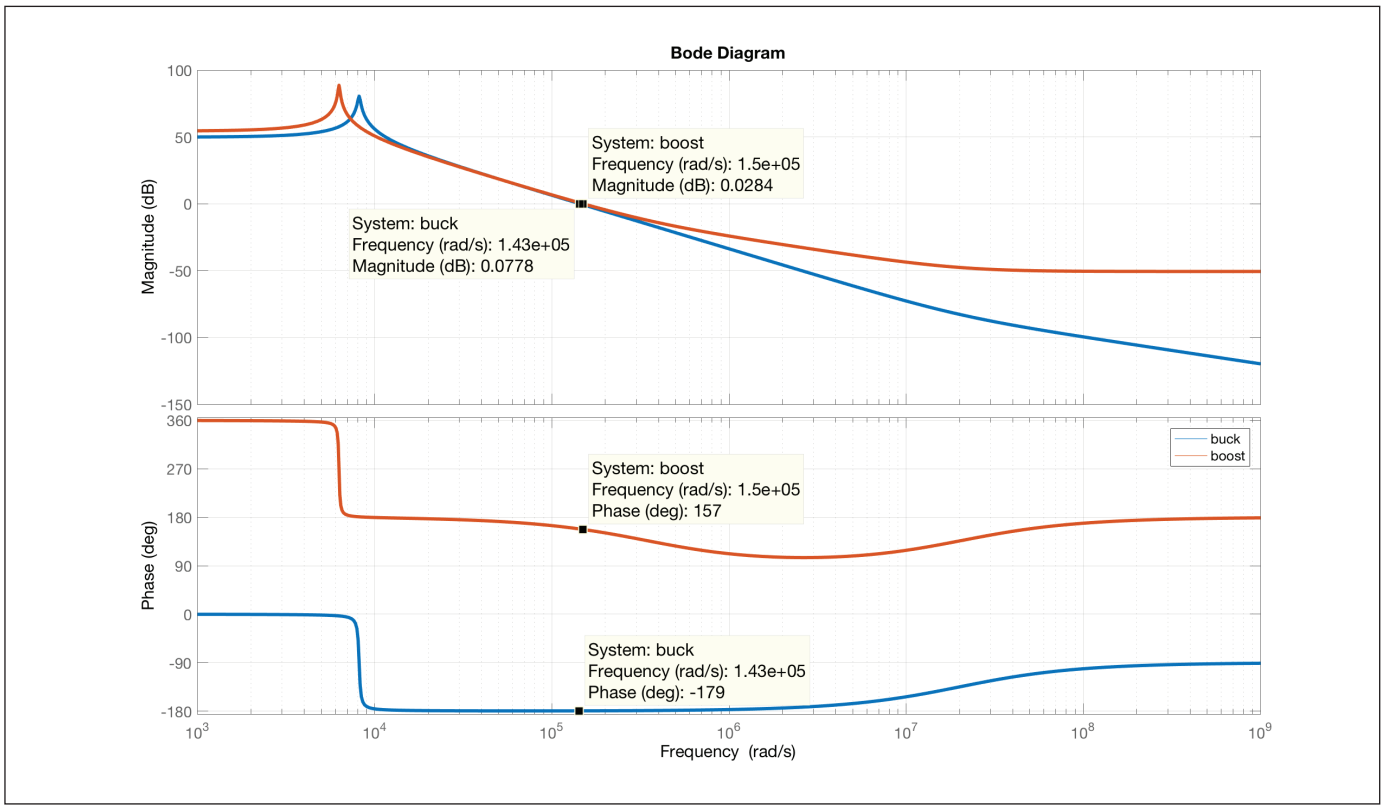

**Figure 5.** Frequency response diagram of the non-inverting buck-boost converter.

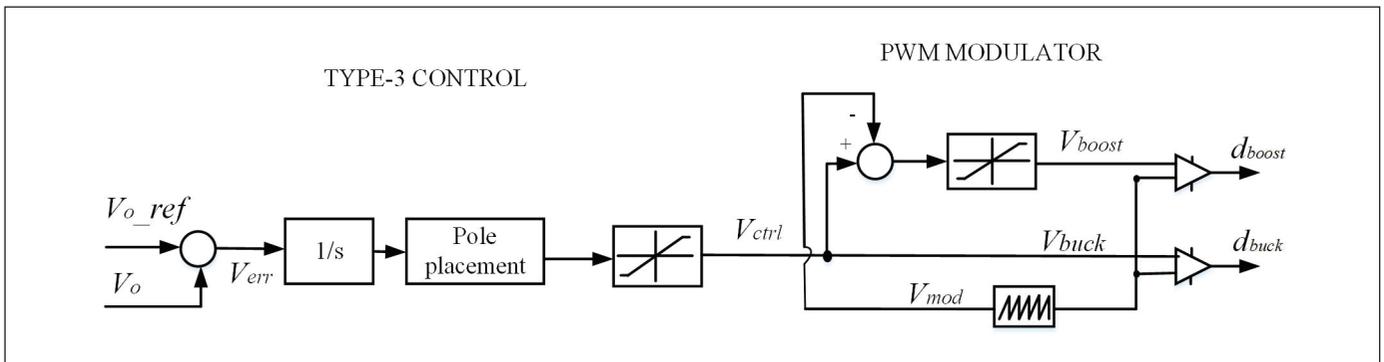

**Figure 6.** Control flow diagram.

The reference 400 V DC value is applied for the boost mode, while the 280 V DC reference value is applied for the buck mode. The tracking performance of the nominal voltage value of 310 V DC reference was also investigated. Firstly, boost mode was initially simulated for the lamp starting and then the converter switched to the buck mode for operation. The obtained simulation results are given in Figure 7. This figure shows that the compansator is able to achieve the performance requirements. It is transparent that the one-level controller in buck mode displays better performance.

The Type-III controller is also synthesized to minimize the unwanted state originating from this feature. The obtained performances from the closed-loop system using the Type-III controller are given in Table 5.

**Table 5.** Performance parameters of the non-inverting buck-boost converter.

| Parameter | Value |
|-----------|-------|
| Rise time [$t_r$] | 0.06 ms |
| Settling time [$t_s$] | 0.20 ms |
| Steady state error [$e_{ss}$] | 0 |
| Overshoot [$M_p$] | % 13 |

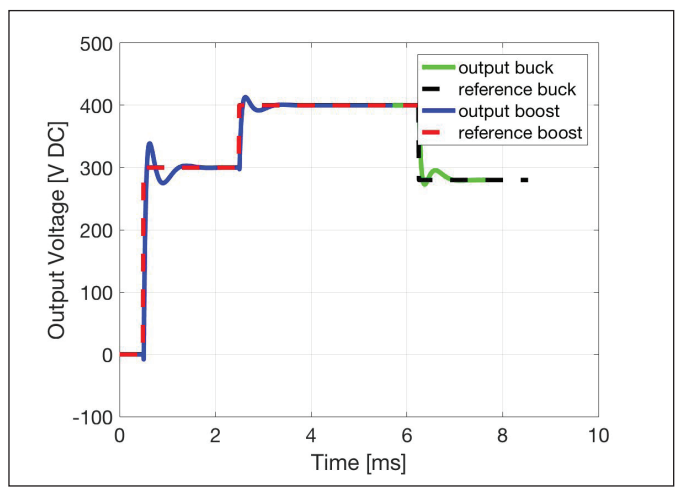

**Figure 7.** Reference tracking performance of converter with Type-III controller.

## 6. Conclusion

Design and control of the cascaded buck boost converter is presented for electronic ballasts, which is compatible with LED drivers. It provides simple solution for igniting fluorescent lamps at starting. Thus, there is no need for frequency changes in the resonance circuit at starting. Also, the circuit can adapt different load voltage levels. A feedback control system, with a type III compensator, is designed to ensure small signal stability with adequate gain and phase margins. The concept has been indicated by evaluating 18 W fluorescent lamp prototype. The simulation results are presented to verify the feasibility and performance of the proposed method.

## 7. Acknowledgment

This research was supported by Projects of Scientific Investigation Unit of Zonguldak Bulent Ecevit University.